\documentstyle[12pt]{article}
\newcommand{\be}{\begin{eqnarray}}
\newcommand{\ee}{\end{eqnarray}}

\newcommand{\bee}{\begin{enumerate}}
\newcommand{\een}{\end{enumerate}}
\newcommand{\bs}{\bigskip}
\begin{document}
\input epsf.tex
\begin{center}
{\Huge\bf  Generalized Background-Field Method}
\end{center}
\bs
\begin{center}
{Y.J. Feng$^*$ and C.S. Lam$^{\dag}$}\\
\bs
{\it Department of Physics, McGill University,\\
3600 University St., Montreal, P.Q., Canada H3A 2T8}
\end{center}

\begin{abstract}

The graphical method  discussed  
previously can be  used to create new
gauges not reachable by the path-integral formalism.
By this means a new gauge is designed for more efficient 
two-loop QCD calculations.  It is related to but simpler than 
the ordinary background-field gauge, in that even the
triple-gluon vertices for internal lines contain only four terms, 
not the usual six. This reduction simplifies the calculation inspite
of the necessity to include other vertices for compensation. 
Like the ordinary background-field gauge, this generalized background-field 
gauge also preserves gauge invariance of the external particles.
As a check of the result and an illustration for the reduction in labour, 
an explicit calculation of the two-loop QCD $\beta$-function
is carried out in this new gauge. It 
results in a saving of 45\% of computation
compared to  the ordinary background-field gauge.
\end{abstract}

\section{Introduction}

Physical processes in QCD are gauge independent but 
unfortunately individual Feynman
diagrams are not. For that reason calculations may be greatly simplified
with the choice of a convenient gauge, so as to 
minimize the presence of gauge-dependent
terms in the intermediate steps. 
The background-field
(BF) gauge \cite{BFM} is  one such  gauge, partly
because of its gauge-invariant property 
with respect to the external lines.
The pinching technique \cite{PT,PTnew} used to simplify calculations 
is also known 
to be related to this gauge \cite{PB,YF1}.

The purpose of this paper is to discuss a graphical method for
designing other convenient gauges. We start by pointing  out the 
advantage and the flexibility of the graphical method over the conventional
path-integral or operator technique.

The gluon propagator $g^{\mu\nu}/p^2$ will be used throughout, thus by a
gauge choice we just mean the choice of vertices in making calculations.
The BF vertices are different from the ordinary vertices in that 
its triple-gluon (3g) vertex makes a distinction between internal
and external gluon lines, with the latter indicated graphically by
an arrow (see Fig.~7(a) of the Appendix) and analytically possessing
only four (see eq.~(9) of the Appendix) rather than the usual
six (eq.~(17) and Fig.~7(i)) terms. The Gervais-Neveu
gauge \cite{GN} would be another possible gauge choice in this sense.

In the usual approach, gauge choice is implemented by a gauge-fixing
term in the path integral. In the BF gauge, for example, this term
for the quantized Yang-Mills field $Q$ is given by $\partial\cdot Q+g[A,Q]$,
with $A$ being an external classical Yang-Mills potential. The presence of
$A$ is the reason why external lines play
a special role in the BF gauge.

In a previous publication \cite{YF1} we have demonstrated how this and
other gauge choices can be obtained in a graphical method,  
which we will summarize and extend in Sec.~2. Similar techniques
have also been employed elsewhere \cite{radial,parti}. 
Essentially, in the graphical language,
the fundamental difference between one gauge and another
lies in their 3g vertices, which differ from one another by a combination
of gradient terms. To compensate for this difference  changes will have to
be made in other vertices as well,  changes that can be computed using the
graphical method. With this technique it is possible  to make
different changes on different vertices, each leading to a different
compensation. In contrast, a gauge-fixing term
in the path-integral  or the operator formalism does not possess this
flexibility; whatever changes made to one vertex must be made on other
identical vertices. Hence we can design newer
and simpler gauges using the graphical method that cannot be obtained
using the path-integral method. The generalized background-field [GBF]
gauge discussed later is such a gauge, and we are also considering
another one that is related to the Gervais-Neveu gauge \cite{YF3}.

This paper is organized as follows. In Sec.~2, we review 
and extend the
graphical procedure for creating new gauges \cite{YF1}. The GBF gauge
will be defined in
Sec.~3, and the computation of the 2-loop $\beta$-function 
using this new gauge 
is presented in Sec.~4, with a conclusion in Sec.~5. 

\section{Graphical Procedure for the Creation of
 New Gauges}

To discuss gauge invariance graphically, it is convenient to use
the Chan-Paton color factors \cite{cp} and color-oriented diagrams
\cite{RG,sr}.

In the absence of quarks, the Chan-Paton 
color factors are given by the
traces of products of the color matrices $T^a$, and the products
of such traces. In the presence of quarks, ordered products of
the color matrices $T^a$ also enter.

Diagrammatic rules can be designed
to compute the spacetime amplitudes for each of these color factors, by
using {\it color-oriented diagrams}. The propagators of the
color-oriented diagrams are the ordinary propagators; their vertices
are different but can be derived from the vertices of Feynman diagrams.
For one thing, color factors are no longer present in the vertices of
the color-oriented diagrams. For another, the
clockwise orientation of the lines emerging from
 each vertex is fixed in the color-oriented
diagrams (hence the name).
These color-oriented vertices are given in eqs~(2.1)
to (2.4) as well as Fig.~1 of Ref.~[9] in the Feynman gauge.

In what follows, when we talk about diagrams we mean
the color-oriented diagrams, and when we talk about
vertices we always refer
to these color-oriented vertices.

The graphical rules for gauge transformation of color-oriented diagrams 
have been discussed before \cite{YF1}. Using these rules we can
create new gauges not reachable in the path-integral formalism.
By new gauges in this paper we shall mean new vertex factors;
the propagators used here will always be the usual Feynman-gauge
propagators.

To create a new gauge $B$ from an existing gauge $A$, we start 
by subtracting appropriate combinations of
gradient terms from the triple-gluon (3g) vertices 
$\Gamma_{\alpha\beta\gamma}(p_1,p_2,p_3)$ of gauge A, 
{\it viz.,} terms proportional to $(p_1)_\alpha, (p_2)_\beta$, and
$(p_3)_\gamma$. Such a gradient term will be denoted graphically
by a cross ($\times$) on the appropriate gluon line.

To maintain gauge invariance and the same physical scattering
amplitudes, other vertices must also be altered and/or created
to compensate for this change. Using graphical methods \cite{YF1}
they can be computed in the following way.

A gradient term on a 3g vertex becomes a divergence on the 
subsequent vertex. To find out the effect of this
change on the 3g vertex, we need to know
the divergence
of every vertex possessing a gluon line in the original gauge $A$. 

If $A$ consists of the vertices of the ordinary Feynman gauge 
given by eqs.~(2.1) to (2.4) 
and Fig.~1 of Ref.~\cite{YF1}, then
these divergences are given by eqs.~(2.5) to (2.8) and expressed
graphically in Figs.~2 to 5 of that paper. 

If $A$ consists of the vertices of the background-field (BF) gauge 
given by eqs.~(2.1) to (2.4), as well as
(4.1) to (4.7)
of Ref.~\cite{YF1},  and graphically Figs.~1 and 18 of that paper,
then these divergences 
can be similarly computed.
The BF vertices are repeated here in the Appendix, eqs.~(7) to (17),
and graphically in Figs.~7(a) to (k). The result of the
divergence computation can be found in  Figs.~8 to 13.

The next step is to combine the right-hand side of the divergence relations
from different diagrams.
On account of local gauge invariance, many of these terms add up to cancel
one another. Starting from the Feynman gauge, these cancellation relations
can be found in Figs.~7 to 16 of Ref.~\cite{YF1}. Starting from the BF gauge,
similar relations can be worked out and they are shown in Figs.~14 to 17 
in the Appendix of this paper.

This general scheme works in QED as well as  QCD. What makes them different
is the presence of ghost lines in the latter. Among other things, it
gives rise to the presence of {\it propagating diagrams} in the divergence
relation of 3g and possibly 4g vertices. 
These are diagrams in which two of the gluon
lines are replaced by ghost lines (we call them {\it wandering ghost} lines),
and the divergence `cross' at the beginning
of a line is moved to the end of the other line, as in
Figs.~3(d) and 3(e) of Ref.~[9] for the Feynman gauge, and
Figs.~8(d) and 8(e) of this paper for the BF gauge. 
Via these diagrams, the divergence `cross' propagates along the diagram,
dragging behind it the wandering ghost line. It is the presence of
these diagrams that makes the Slavnov-Taylor identity in QCD different
from the Ward-Takahashi identity in QED.

If the cross propagates in a closed loop to return to its original position,
then local gauge compensation will be upset, thus resulting in additional terms
or vertices with two ghost lines. For example,
if gauge $A$ is Feynman and gauge $B$ is BF, then this change
is given by Fig.~20 of Ref.~\cite{YF1}.

We started with the gradient change of a {\it single} 3g vertex
in gauge $A$ for every diagram, as described above, and
considered the local cancellations and 
the new ghost vertices thus created for gauge $B$.
Having converted one vertex this way, 
we are now ready to convert a {\it second} 3g
vertex from gauge $A$ to gauge $B$. Unless the second vertex is adjacent
to the first, the same argument holds and the net change of the
second vertex is identical
to that of the first. If they are adjacent, additional changes
may occur because the first 3g and ghost vertices 
are already in gauge $B$, while the second vertex is still in gauge $A$.
Mixing $A$ and $B$, 
local gauge cancellation will generally not occur, thus producing yet other
new terms or vertices from this mismatch. 
In getting from the Feynman gauge to the
BF gauge in Ref.~\cite{YF1}, for example,
this is how the new 4g vertices are obtained through
Figs.~37 and 38.

We can continue this way to change {\it all other} vertices 
one after another from gauge $A$
to gauge $B$. In principle, merging three or more adjacent vertices
may produce newer vertices still, 
but this does not happen when we go from the
Feynman to the BF gauge, nor from the BF gauge to the GBF gauge as will
be discussed in the next section.

\section{Generalized Background-Field (GBF) Gauge}

In this section we follow the outline of the last section to convert the 
BF gauge to a new gauge which we shall refer to as the GBF gauge. 

In the BF gauge, the 3g vertices involving 
an {\it external} gluon are  different from those without it.
The former has four terms compared to the latter with six terms. 
The former is shown in Fig.~7(a) of the Appendix, in which the external line
is represented by an arrow. The latter is just the usual 3g vertex 
 shown in Fig.~7(i).
The aim of the GFB gauge is to convert {\it all} 3g vertices
to the former type with only four terms, so as to simplify the
number of terms present and the
algebra of the calculations. It is true that new vertices and diagrams will
have to be produced to compensate for this change, 
but the overall saving turns out to be still substantial. In this paper,
we confine ourselves
to two-loop diagrams with an arbitrary number of external lines.
In fact, we will carry out the explicit calculation only for two
external lines but the technique can easily be generalized to an arbitrary
number of external lines.
With this restriction
there are at most two internal 3g vertices in the BF gauge that need
be converted into the arrowed type.
We will choose the arrows to appear on both ends of the `middle propagator'
as shown in Figs.~6(a) and 6(b) of the next section.

As before, let us make this change first on one of the two
internal vertices.
The change is identical to what happens when we convert from the Feynman
to the BF gauge, so nothing new will be produced. Now we convert the second
internal 3g vertex to the arrowed form. Via a series of propagating diagrams,
the `cross' can return to this vertex via two possible routes.
Either it comes back via a line without an arrow, or it returns via the
arrowed line. The former is identical to what happened before so it
produces nothing new. The latter could not happen
previously
 because the arrowed line in the BF gauge is always an external line. 
Now in the GBF gauge, this new situation produces a new vertex 
shown in Fig.~1.
In addition, adjacent interaction may now take place when the ghost line
returns in a loop, and this produces further changes given by
Figs.~2 to 4. 
\begin{figure}
\vskip -0 cm
\centerline{\epsfxsize 4.7 truein \epsfbox {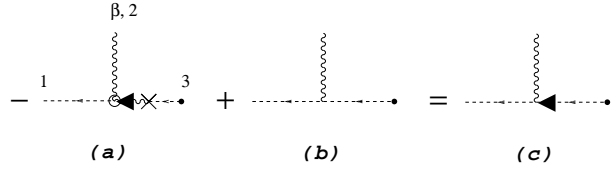}}
\nobreak
\vskip -13cm\nobreak
\vskip .1cm
\caption{
Generation of a ghost vertex in the GBF gauge. The small circle
in (a) represents a factor of $g_{\beta\gamma}$, and the cross denotes
$p_3^{\gamma}$. A dot at the end of a line means that the propagator
of the line is included. The propagator of line-3 in (a) is a ghost propagator. 
}
\end{figure}

\begin{figure}
\vskip -1 cm
\centerline{\epsfxsize 4.7 truein \epsfbox {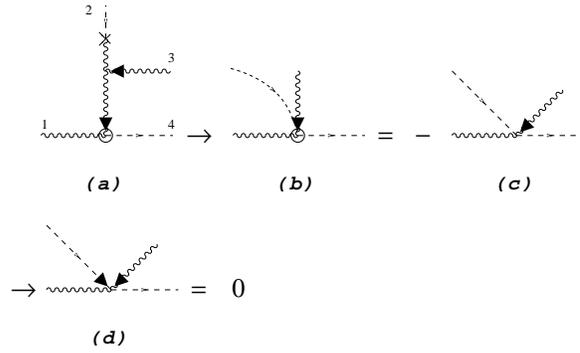}}
\nobreak
\vskip -11cm\nobreak
\vskip .1cm
\caption{Cancellation of a ghost vertex in the GBF gauge. The cross on line-2 
of (a) generates a sliding diagram as shown in (b), which up to a sign is
equal to (c), a ghost coupling already exists
in the BF gauge. (d) is the summation of (b) and (c). It is zero.}
\end{figure}

\begin{figure}
\vskip -0 cm
\centerline{\epsfxsize 4.7 truein \epsfbox {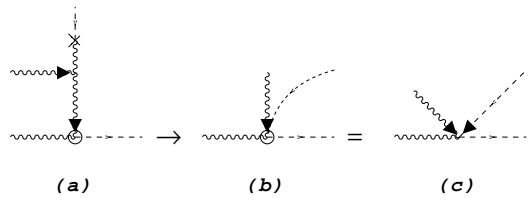}}
\nobreak
\vskip -13cm\nobreak
\vskip .1cm
\caption{Generation of another ghost vertex in the GBF gauge.  }
\end{figure}

\begin{figure}
\vskip -1 cm
\centerline{\epsfxsize 4.7 truein \epsfbox {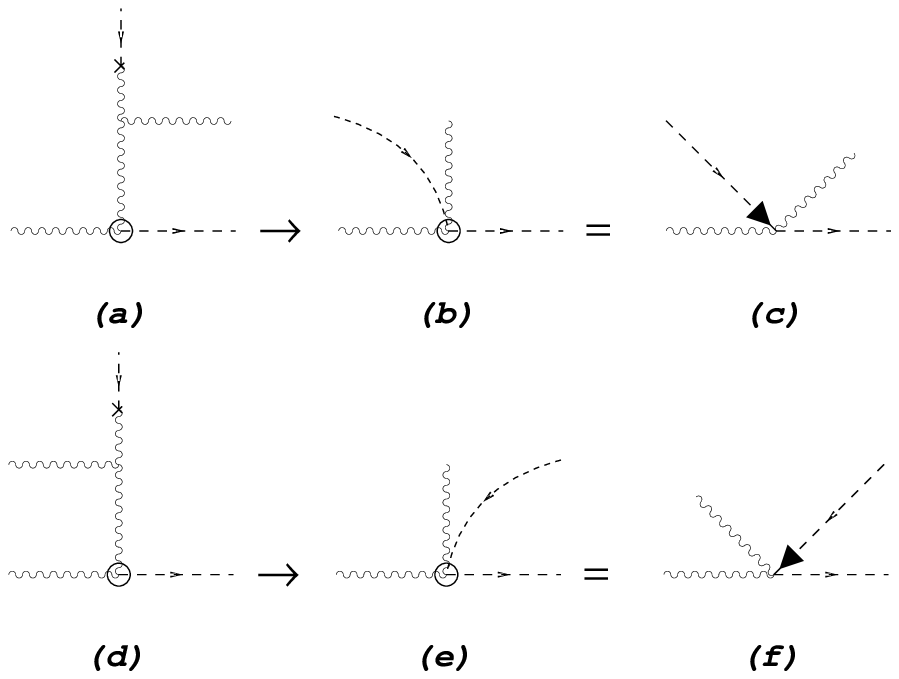}}
\nobreak
\vskip -11cm\nobreak
\vskip .1cm
\caption{Generation of other ghost vertices in the GBF gauge.}
\end{figure}

The vertices obtained this way for the GBF gauge is summarized in Fig.~5
and eqs.~(1)--(6).
\begin{figure}
\vskip -2.5 cm
\centerline{\epsfxsize 4.7 truein \epsfbox {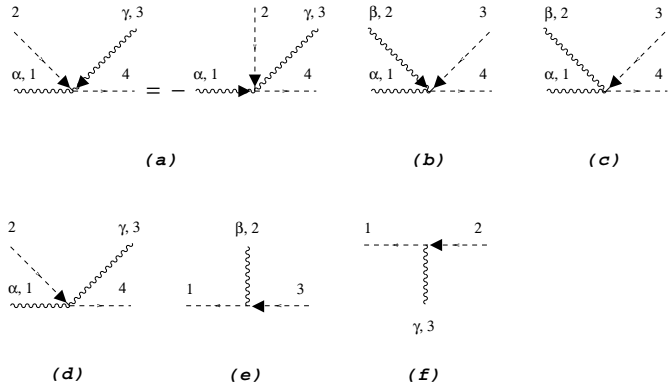}}
\nobreak
\vskip -11cm\nobreak
\vskip .1cm
\caption{New vertices in GBF gauge. Vertices involving the arrowed ghost lines
originates from the internal arrowed gluon lines in the GBF gauge. All the BF gauge vertices in Fig.~7 are still present.}
\end{figure}

\begin{eqnarray}
&&[5a]=0  \\
&&[5b]=g^2 g^{\alpha\beta} \\
&&[5c]=g^2 g^{\alpha\beta} \\
&&[5d]=g^2 g^{\alpha\gamma} \\
&&[5e]=-g p_2^\beta \\
&&[5f]=g p_3^\gamma
\end{eqnarray}

\section{$\beta$-Function at the Two-Loop level}

As an illustration, and a check of these new vertex rules, we will calculate the QCD two-loop $\beta$ 
function in the GFB gauge. This $\beta$-function 
has previously been done in the Feynman and 
the BF gauges \cite{BFM2}, with considerable savings shown when computed
in the BF gauge. We will now show that a further saving of 45\%
is possible when computed in the GBF gauge.

We choose this example for illustration because it
 is the simplest at the two-loop level 
and it can be computed analytically.
The disadvantage of this example is that it is not an on-shell process, so the
full-fledged simplification of the GBF gauge will not be revealed. Off-shellness
gives rise to some {\it extra} diagrams, which will be absent in 
an on-shell process. But even so, the saving is still considerable.

The diagrams in the GBF gauge are shown in Fig.~6. There are 26
basic diagrams to be found in Fig.~6(a--z),  
12 {\it extra} diagrams to be found in Fig.~6(e1--e12), 
and 3 gauge-fixed renormalization insertion diagrams to be found
in Fig.~6(i1, i2, i3). 
The extra diagrams involve wandering ghost lines sliding
to an external end. They will be absent if these external lines were
on-shell.

The result of the calculation is summarized in the Table I.

We can compare this with the calculation in the BF gauge \cite{BFM2}. 
Although  we have 
more diagrams in the GBF gauge,  the total number of terms 
to be computed is less.
In the GBF gauge, there are totally 728 terms, while in the BF gauge there are 
1320 terms. Therefore, a 45\% 
of computational labor is saved by using the GBF gauge.
Using $Mathematica^R$ to compute, we need 150 seconds for the GBF gauge, and
260 for the BF gauge.
For on-shell process, because of the absence of the extra diagrams, 
the saving of the GBF gauge will be greater and
can be expected to be approximately 50 percent.

\begin{center} 
\begin{tabular}{|c|c|c|c|c|}   \hline
Graph  &\multicolumn{2}{c|} {$g^{\mu\nu} p^2 $}   & \multicolumn{2}{c|} 
         {$p^\mu p^\nu $}  \\ \cline{2-5}
       &${1\over\epsilon^2}$&${1\over\epsilon}$&${1\over\epsilon^2}$&${1\over \epsilon}$
                                             \\ \hline
a      & $ 11/3$ & $  (54-22\rho)/3$
       & $-11/3$ & $ -(56-22\rho)/3$ \\ \hline
b      & $ 10/3$ & $  (52-20\rho)/3$
       & $-10/3$ & $ -(53-20\rho)/3$ \\ \hline
c      & $-1   $ & $  0            $
       & $ 0   $ & $  0            $ \\ \hline
d, p   & $-1/2 $ & $  -(3-2\rho)/2 $ 
       & $ 1/2 $ & $ (5-2\rho) /2  $ \\ \hline
e, k   & $ 5/24$ & $ (65-20\rho)/48$
       & $-5/24$ & $-(33-10\rho)/24$ \\ \hline
f,j    & $-1/8 $ & $  -(9-4\rho)/16$
       & $ 1/8 $ & $   (9-4\rho)/16$ \\ \hline
g,m    & $ 1/6 $ & $  (13-4\rho)/12$
       & $-1/6 $ & $   -(3-\rho)/3 $ \\ \hline
h,i    & $-5/48$ & $(-41+20\rho)/96$
       & $ 5/48$ & $ (45-20\rho)/96$ \\ \hline
l      & $  1/6$ & $  (19-4\rho)/12$
       & $ -1/6$ & $ -(15-4\rho)/12$ \\ \hline
n,o    & $ 1/24$ & $  (13-4\rho)/48$
       & $-1/24$ & $   -(3-\rho)/12$ \\ \hline
q      & $    0$ & $            1/2$ 
       & $    0$ & $              0$ \\ \hline
r,s,t,u& $ -1/8$ & $ -(13-4\rho)/16$
       & $  1/8$ & $   (9-4\rho)/16$ \\ \hline
v,w    & $-1/48$ & $ -(13-4\rho)/96$
       & $ 1/48$ & $   (9-4\rho)/96$ \\ \hline
x,y    & $    0$ & $              0$
       & $    0$ & $              0$ \\ \hline
z      & $   -6$ & $   -(24-12\rho)$
       & $    6$ & $    (24-12\rho)$ \\ \hline
e1     & $  5/8$ & $ (33-20\rho)/16$
       & $    0$ & $              0$ \\ \hline
e2     & $-9/16$ & $-(217-36\rho)/32$
       & $    0$ & $             9/2$ \\ \hline
e3     & $  1/2$ & $     (5-2\rho)/2$
       & $    0$ & $               0$ \\ \hline
e4     & $-9/16$ & $ -(73-36\rho)/32$
       & $    0$ & $               0$ \\ \hline
e5     & $    0$ & $            -1/8$  
       & $    0$ & $               0$ \\ \hline
e6     & $ -1/8$ & $   -(9-4\rho)/16$ 
       & $    0$ & $               0$ \\ \hline
e7     & $ -1/8$ & $  -(25-4\rho)/16$
       & $    0$ & $             1/4$ \\ \hline
e8     & $  1/4$ & $     (5-2\rho)/4$
       & $    0$ & $               0$ \\ \hline
e9     & $ -1/8$ & $  -(13-4\rho)/16$
       & $    0$ & $               0$ \\ \hline
e10    & $-3/16$ & $ -(27-12\rho)/32$
       & $    0$ & $               0$ \\ \hline
e11    & $-3/16$ & $ -(75-12\rho)/32$
       & $    0$ & $             3/4$ \\ \hline
e12    & $  1/2$ & $     (5-2\rho)/2$
       & $    0$ & $               0$ \\ \hline
i1     & $ 25/9$ & $ (230-75\rho)/27$
       & $-25/9$ & $-(230-75\rho)/27$ \\ \hline
i2,i3  & $-25/18$& $-(140-75\rho)/54$
       & $ 25/18$& $ (140-75\rho)/54$ \\ \hline
Total  & $    0$ & $            17/3$
       & $    0$ & $           -17/3$\\ \hline
\end{tabular}
\ 

{Table I: The result of two-loop $\beta$-function in the GBF gauge. $\rho$ is 
the same parameter as used in Ref.~13.}
\end{center}

\begin{figure}
\vskip -0 cm
\centerline{\epsfxsize 4.7 truein \epsfbox {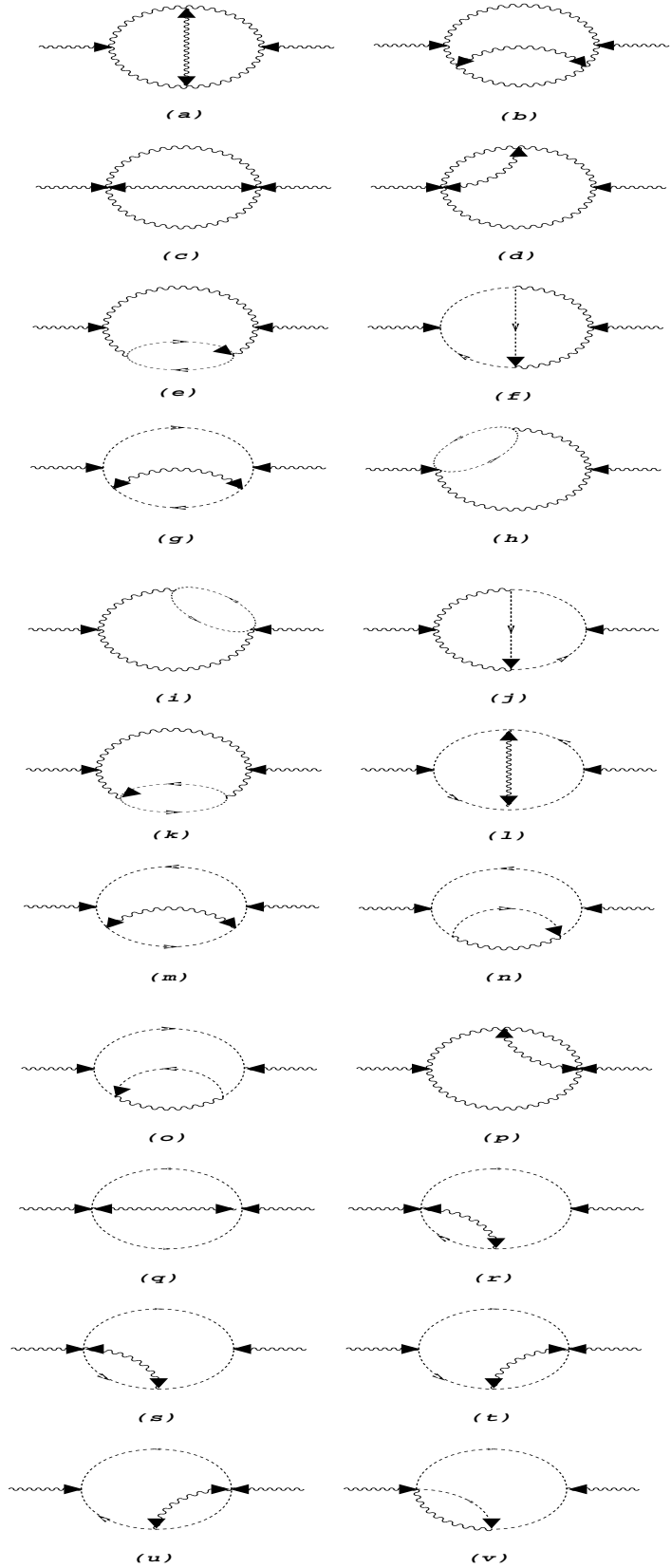}}
\nobreak
\vskip 5 cm\nobreak
\vskip .1cm
\caption{Feynman diagrams in GBF gauge for two-loop $\beta$-function calculation of QCD.}
\end{figure}
  
\setcounter{figure}{5}

\begin{figure}
\vskip -0 cm
\centerline{\epsfxsize 4.7 truein \epsfbox {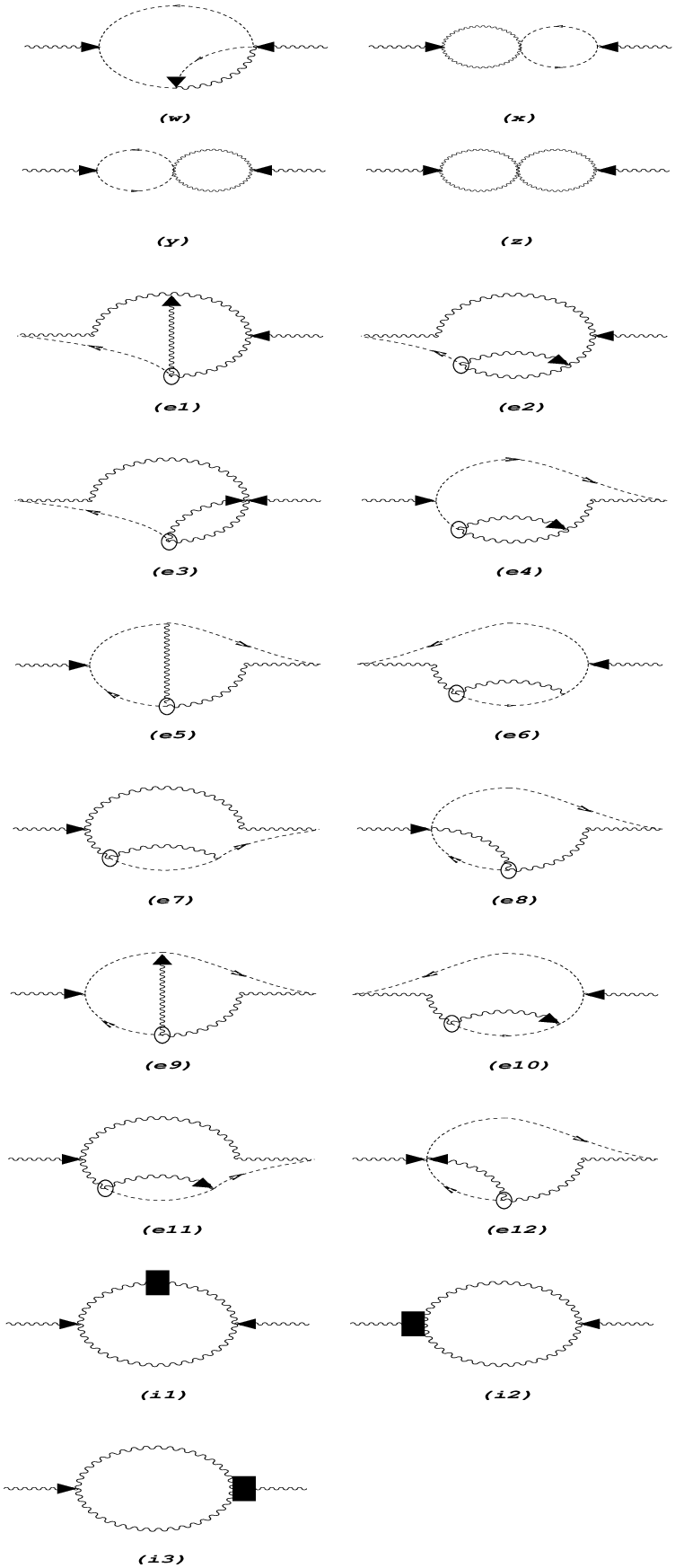}}
\nobreak
\vskip 5 cm\nobreak
\vskip .1cm
\caption{(continued)}
\end{figure}

\section{Conclusion}
We have demonstrated the power of the graphical rules in
making individual gauge changes on vertices.
An operator method or a path-integral method must treat all the
vertices the same way so this 
individual flexibility is lost.
This method is illustrated by the
creation of the GBF gauge from the BF gauge. 
GBF gauge maintains 
the gauge-invariant property of the ordinary BF gauge 
with respect to the external lines, and preserves the simple
Ward-Takahashi identity when divergences are taken on
them. It also contains less terms than the BF gauge
in actual calculations. 
The saving for the two-loop QCD $\beta$-function we gave is  45\%,
and more can be expected for on-shell processes.

The graphical method is not limited to this example
nor this gauge. We can for example change the 
internal $3g$ vertices into the $3g$ vertices of the 
Gervais-Neveu gauge \cite{GN}, 
with even greater saving \cite{YF3}.

\section{Acknowledgements}
This research was supported in part by the Natural Science and Engineering 
Research Council of Canada and by the Qu\'{e}bec Department of Education. 
Y.J.F. acknowledges the support of the Carl Reinhardt Major Foundation.

\appendix
\section{Divergence and cancellation relations in the BF gauge}
The color-oriented vertices 
of QCD in the BF gauge are summarized in Fig.~7. 
We use wavy line for gluon, dotted
line for ghost. The arrowed line is an external 
line. All propagators in this paper 
are chosen to be in the Feynman gauge, so  we
have $-1/p^2$ and $g^{\alpha\beta}/p^2$ for ghosts and gluons respectively.

\begin{figure}
\vskip -0 cm
\centerline{\epsfxsize 4.7 truein \epsfbox {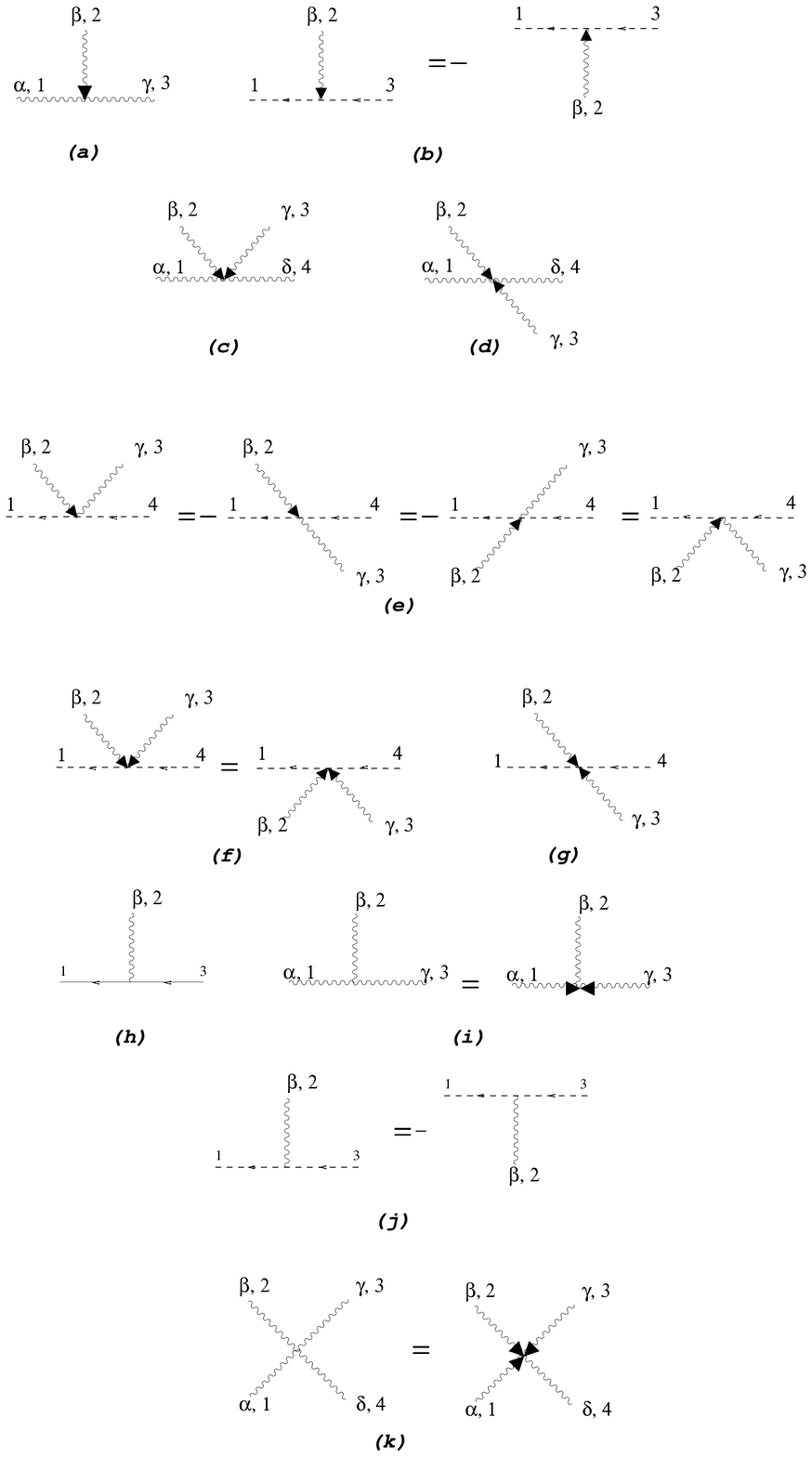}}
\nobreak
\vskip -2cm\nobreak
\vskip .1cm
\caption{Color-oriented vertices of QCD in the BF gauge.}
\end{figure}

Analytically, the vertices shown in Fig.~7 are associated with the following
vertex factors

\begin{eqnarray}
&&[7a]=g\ [g_{\gamma\alpha}(p_3-p_1)_\beta-2g_{\alpha\beta}(p_2)_\gamma
+2g_{\beta\gamma}(p_2)_\alpha]\ ,\\
&&[7b]=g\ (p_1-p_3)_\beta\ ,\\
&&[7c]=g^2\ [-g_{\beta\gamma}g_{\alpha\delta}-2g_{\alpha\beta}g_{\gamma\delta}
+2g_{\alpha\gamma}g_{\beta\delta}] ,\\
&&[7d]=-2g^2\ g_{\beta\gamma}g_{\alpha\delta}\ ,\\
&&[7e]=g^2\ g_{\beta\gamma}\ ,\\
&&[7f]=g^2\ g_{\beta\gamma}\ ,\\
&&[7g]=-2g^2\ g_{\beta\gamma}\ , \\
&&[7h]=g\  \gamma_{\beta} \ , \\
&&[7i]=g[ g_{\alpha\beta} (p_1-p_2)_{\gamma}+
g_{\beta\gamma}(p_2-p_3)_{\alpha}
+g_{\gamma\alpha}(p_3-p_1)_{\beta}] \ , \\
&&[7j]=g\  (p_1)_{\beta}  \ , \\
&&[7k]=g^2\  \left[ 2 g_{\alpha\gamma}
g_{\beta\delta}-g_{\alpha\beta}g_{\gamma\delta}-g_{\alpha\delta}
g_{\beta\gamma}\right] \ .
\end{eqnarray}

The divergence and cancellation relations for the
BF gauge are summerized below. 

\subsection{Divergence relations}
We list below the possibilities when a cross is put on a
gluon line without an arrow for any of the vertices found in Fig.~7.

\begin{enumerate}
\item $3g$ vertex:
this is shown in Fig.~8. 
\begin{figure}
\vskip -1 cm
\centerline{\epsfxsize 4.7 truein \epsfbox {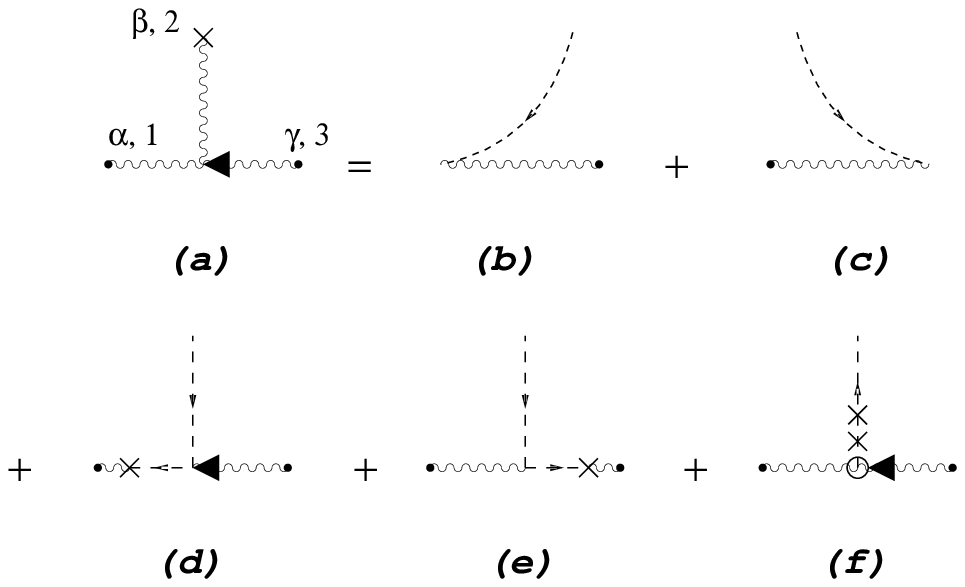}}
\nobreak
\vskip -11cm\nobreak
\vskip .1cm
\caption{Divergence relation of the triple-gluon vertex (in the BF gauge). 
(b) and (c) are sliding diagrams. The wandering ghost line tangential to the gluon line means 
the corresponding gluon propagator has been cancelled. The vertex factor
for (b) and (c) is just $g g_{\alpha\gamma}$, and the sign convention is
that if the wandering ghost line turns to its right with respect to its
ongoing direction, like (b), the sliding diagram carries a plus sign. If it turns to its left like (c), the diagram carries a minus sign. In (f), two 
crosses  are head to head on line-2. So the propagator of line-2 is cancelled.}
\end{figure}

\item $4g$ vertex:
\begin{enumerate} 
\item
$4g$ vertex with one external line:
if the crossed line is diagonal to the
arrowed line, it is shown in Fig.~9. 
If the arrowed line is adjacent to the crossed line, it is shown
in Fig.~10.

\begin{figure}
\vskip - 0 cm
\centerline{\epsfxsize 4.7 truein \epsfbox {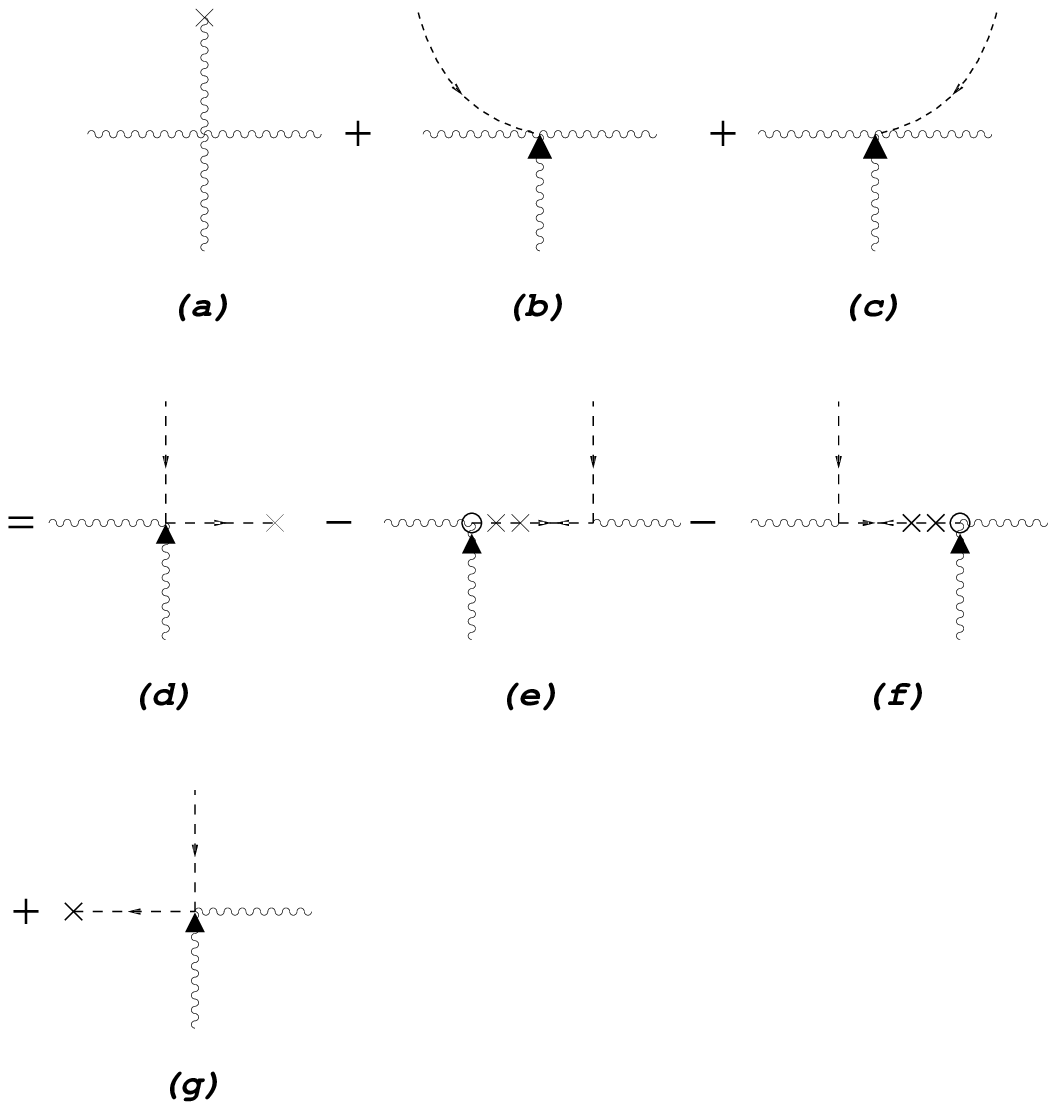}}
\nobreak
\vskip -8.5 cm\nobreak
\vskip .1cm
\caption{Divergence relation of a four-gluon vertex.}
\end{figure}

\begin{figure}[h]
\vskip -0 cm
\centerline{\epsfxsize 4.7 truein \epsfbox {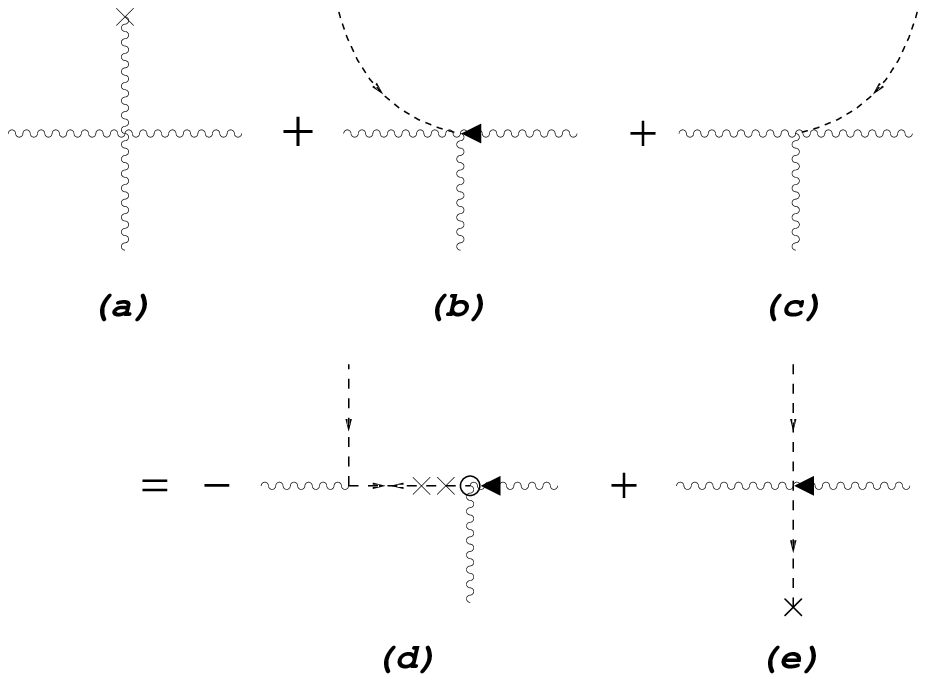}}
\nobreak
\vskip -11.5cm\nobreak
\vskip .1cm
\caption{Another divergence relation of a four-gluon vertex.}
\end{figure}

\item $4g$ vertex with two diagonal external lines: it is shown in Fig.~11.
\begin{figure}[h]
\vskip -.5 cm
\centerline{\epsfxsize 4.7 truein \epsfbox {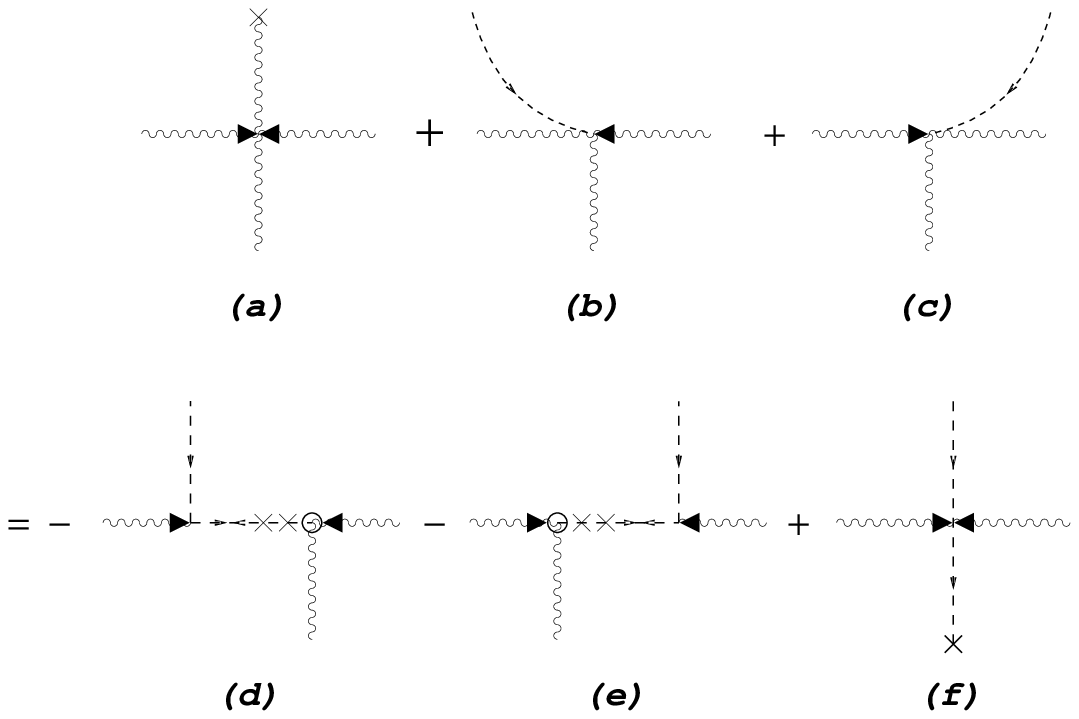}}
\nobreak
\vskip -10.5cm\nobreak
\vskip .1cm
\caption{Divergence relation of a four-gluon vertex with arrowed lines.}
\end{figure}

\item $4g$ vertex with two adjacent external lines:
it is shown in Fig.~12.
\begin{figure}[h]
\vskip -0 cm
\centerline{\epsfxsize 4.7 truein \epsfbox {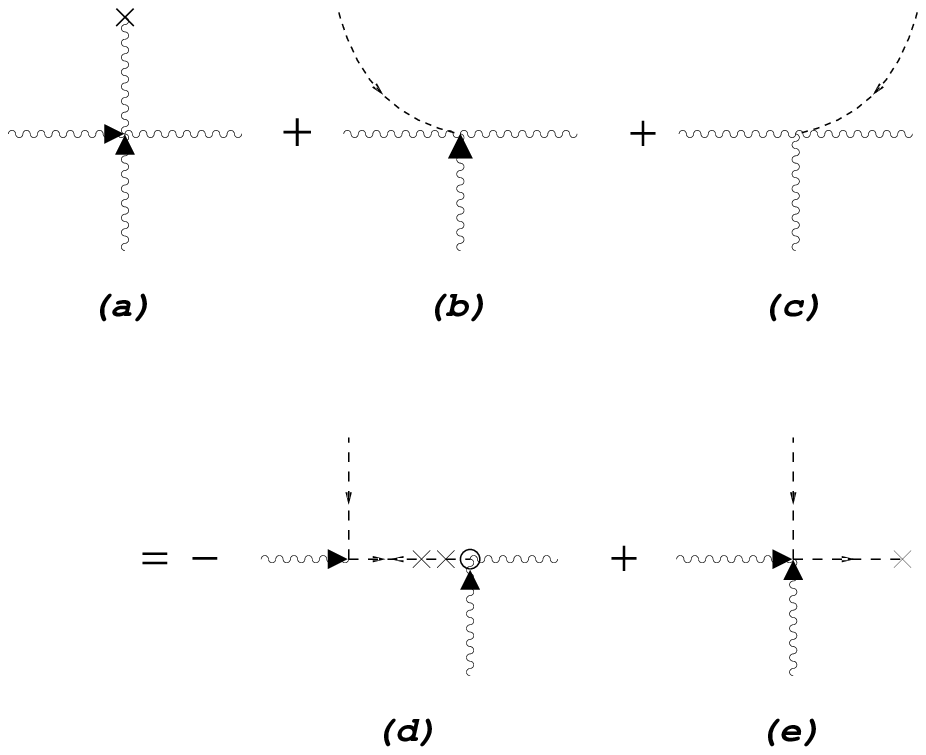}}
\nobreak
\vskip -10cm\nobreak
\vskip .1cm
\caption{Another divergence relation of a four-gluon vertex with arrowed lines.}
\end{figure}
\end{enumerate}

\item Ghost vertex with one external gluon and one internal gluon:
it is shown in Fig.~13.

\begin{figure}[h]
\vskip -.5 cm
\centerline{\epsfxsize 4.7 truein \epsfbox {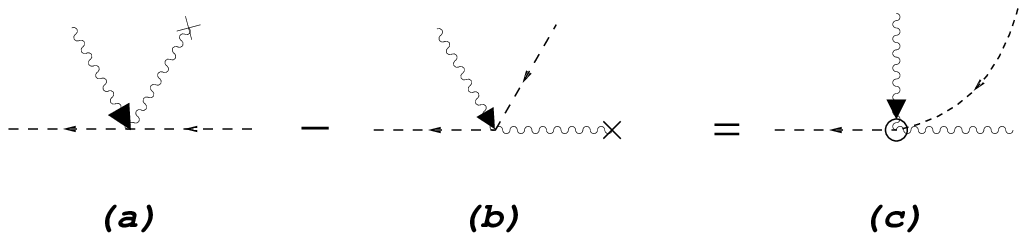}}
\nobreak
\vskip -14cm\nobreak
\vskip .1cm
\caption{Divergence relation of a ghost vertex.}
\end{figure}

\end{enumerate}

\subsection{Cancellation relations}
\begin{enumerate}
\item Cancellation relation involving ghost vertices with one external
and one internal gluon line is
 shown in Fig.~14.

\begin{figure}[h]
\vskip -0.5 cm
\centerline{\epsfxsize 4.7 truein \epsfbox {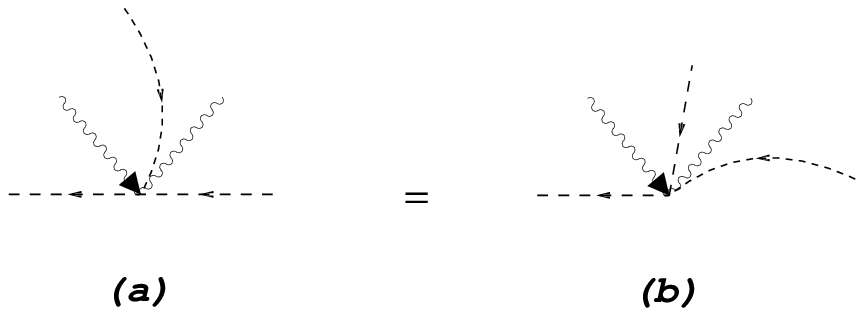}}
\nobreak
\vskip -14cm\nobreak
\vskip .1cm
\caption{An identity (in the BF gauge).}
\end{figure}

\item Cancellation relation involving the ghost vertex with two
 external gluon
lines is
 shown in Fig.~15. 

\begin{figure}[h]
\vskip -.5 cm
\centerline{\epsfxsize 4.7 truein \epsfbox {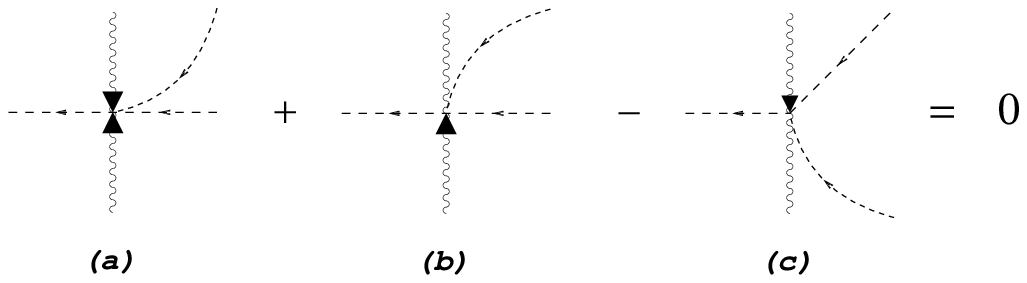}}
\nobreak
\vskip -13.5cm\nobreak
\vskip .1cm
\caption{A cancellation relation.}
\end{figure}

\item Cancellation relation involving
the four gluon vertex with two diagonal external 
lines is shown in Fig.~16.
\begin{figure}[h]
\vskip -1 cm
\centerline{\epsfxsize 4.7 truein \epsfbox {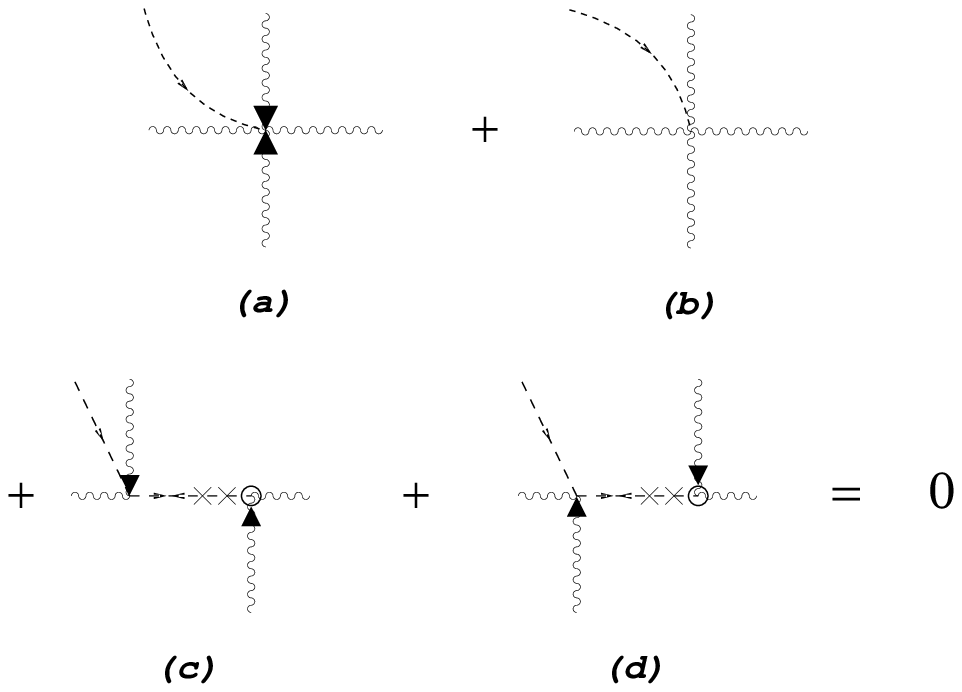}}
\nobreak
\vskip -10cm\nobreak
\vskip .1cm
\caption{Another cancellation relation.}
\end{figure}

\item Cancellation relation involving the
 four gluon vertex with two adjacent external 
lines is shown in Fig.~17.

\begin{figure}[h]
\vskip -0 cm
\centerline{\epsfxsize 4.7 truein \epsfbox {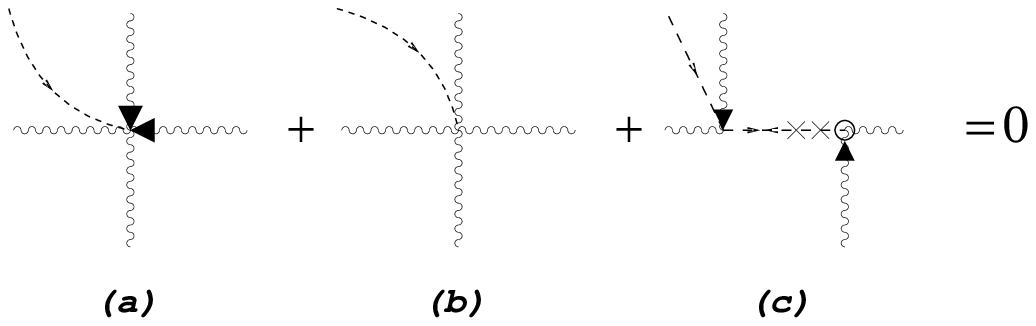}}
\nobreak
\vskip -14cm\nobreak
\vskip .1cm
\caption{Yet another cancellation relation.}
\end{figure}

\end{enumerate}

\end{document}